\documentclass{ws-ijmpa}
\usepackage[super,compress]{cite}
\usepackage{graphicx}
\usepackage{hyperref}
\begin{document}
\markboth{Prabal Adhikari, Jaehong Choi}{Magnetic Vortices in the Abelian Higgs Model with Derivative Interactions}

%
\catchline{}{}{}{}{}
%

\title{Magnetic Vortices in the Abelian Higgs Model with Derivative Interactions}

\author{PRABAL ADHIKARI}

\address{Physics Department, Faculty of Natural Sciences and Mathematics, St. Olaf College, 1520 St. Olaf Avenue\\
Northfield, MN 55057, United States of America\\
adhika1@stolaf.edu}

\author{JAEHONG CHOI\footnote{graduated May 2018}}

\address{St. Olaf College, 1520 St. Olaf Avenue,\\
Northfield, MN 55057, United States of America\\
choi1@stolaf.edu}

\maketitle


\begin{abstract}
We study the properties of a single magnetic vortex and magnetic vortex lattices in a generalization of the Abelian Higgs model containing the simplest derivative interaction that preserves the $U(1)$ gauge symmetry of the original model. The paper is motivated by the study of finite isospin chiral perturbation theory in a uniform, external : since pions are Goldstone bosons of QCD (due to chiral symmetry breaking by the QCD vacuum), they interact through momentum dependent terms. We introduce a uniform external magnetic field and find the asymptotic properties of single vortex solutions and compare them to the well-known solutions of the standard Abelian Higgs Model. Furthermore, we study the vortex lattice solutions near the upper critical field using the method of ``successive approximations", which was originally used by Abrikosov in his seminal paper on type-II superconductors. We find the vortex lattice structure, which remains hexagonal as in the standard Abelian Higgs model, and condensation energy of the vortex lattices relative to the normal vacuum (in a uniform magnetic field).
\end{abstract}



\section{Introduction}
Magnetic vortices are topological objects that quantize magnetic flux. They have been studied extensively since type-II superconductors and vortices were first proposed within the context of Ginzburg-Landau (GL) theory for metals by Abrikosov~\cite{Abrikosov:1956sx, abrikosov1957magnetic,abrikosov1988fundamentals}. Their relevance and study have not been limited to metals but extended to particle physics including the Abelian Higgs model~\cite{Nielsen:1973cs}, electroweak theory~\cite{Ambjorn:1988th} and quantum chromodynamics (QCD)~\cite{Chernodub:2014rya}. Since GL theory and the Higgs model are isomorphic~\cite{Harrington:1975mv}, this is not entirely surprising. However, there are significant qualitative differences between the standard GL theory (or the Abelian Higgs model) and electroweak theory (or QCD at large magnetic fields). While magnetic vortices are formed below a certain critical magnetic field, the opposite is true in the context of electroweak theory, where the electroweak vacuum becomes unstable to $W$-boson vortices above a critical magnetic field. Furthermore, it has also been suggested that the QCD vacuum may be populated by a color magnetic flux tube (also called the ``Copenhagen vacuum")~\cite{Ambjorn:1979xi,Ambjorn:1980ms}.

Currently, there is strong interest in the role of magnetic fields in systems described by QCD due to their relevance to magnetars~\cite{Harding:2006qn}, RHIC collisions~\cite{Kharzeev:2004ey,Kharzeev:2007jp,Fukushima:2008xe}, which may lead to the formation of quark-gluon plasma, the study of which is also relevant to the early formative stages of the universe after the Big Bang. Remarkably, it was found that the QCD vacuum at high enough magnetic fields leads to the condensation of $\rho$-mesons~\cite{Chernodub:2011gs,Chernodub:2012bj,Chernodub:2012bq,Chernodub:2014rya} (similar to the electroweak theory), which form a vortex lattice that consists not only of charged mesons ($\rho_{\pm}$) but also the neutral meson ($\rho_{0}$). The possibility of neutral particle condensation was also explored in the context of a simple model, the ``extended Abelian Higgs model" containing both complex and real scalar fields: it was show that the electromagnetically neutral, real fields when coupled to charged fields can potentially condense in a vortex phase through the coupling of the neutral field with the charged fields. Here, motivated by the presence of derivative interactions within $\chi$PT, we construct a simple generalization of the Abelian Higgs model to include derivative interactions and study its implications. 

Additionally, there is also interest in vortices in the context of finite isospin chiral perturbation theory ($\chi$PT), which is the low-energy description of QCD with pions ($\pi_{\pm}$, $\pi_{0}$) as the relevant degrees of freedom. Son and Stephanov~\cite{Son:2000by,Son:2000xc} found that pions at finite isospin form a superfluid. Since the superfluid is electromagnetically charged, unsurprisingly they exhibit superconducting properties in the presence of an external magnetic field. It was shown in Ref.~\cite{Adhikari:2015wva} that pions condense into vortices with no neutral pion condensation ($\pi_{0}$) though this possibility isn't fully ruled out except in single vortices. However, Ref.~\cite{Adhikari:2015wva} did not study the structure of the vortex lattice or condensation energy as a consequence of the derivative interactions of the Goldstone modes (i.e. pions) of QCD. The objective of this paper is to take steps to fill this intellectual gap by studying a generalization of the Abelian Higgs model including the simplest momentum-dependent, i.e. derivative interactions. In subsequent work, we will apply some of the results of this work to study vortex lattices in finite isospin chiral perturbation theory~\cite{Adhikari:new}.

The paper is organized as follows: we begin in Section~\ref{lag} with the Lagrangian of the Abelian Higgs model with an additional derivative interaction that preserves the local $U(1)$ symmetry of the Abelian Higgs Model. We then study the asymptotic properties of single magnetic vortices in Section~\ref{mv1} and generate numerical solutions, followed by the study of vortex lattices and their condensation energy in Section~\ref{condensation}. Finally, in Section~\ref{conclusion}, we present some concluding remarks. 

\section{Lagrangian}
\label{lag}
We begin with the Lagrangian for the Abelian Higgs Model with an additional derivative interaction
\begin{equation}
\begin{split}
\mathcal{L}&=-\frac{1}{4}F_{\mu\nu}F^{\mu\nu}+(D_{\mu}\phi)^{\dagger}(D^{\mu}\phi)-V(|\phi|)+\ell^{2}(D_{\mu}\phi)^{\dagger}(D^{\mu}\phi)\phi^{\dagger}\phi\\
V(|\phi|)&=-m^{2}|\phi|^{2}+\frac{\lambda}{2}
|\phi|^{4}\ ,
\end{split}
\end{equation}
where $D_{\mu}\equiv\partial_{\mu}+ieA_{\mu}$. This convention for the covariant derivative assumes that the field $\phi$ has charge $+e$ which can be assumed to be positive with no loss of generality. Furthermore, the Lagrangian also consists of a derivative interaction term $(D_{\mu}\phi)^{\dagger}(D^{\mu}\phi)\phi^{\dagger}\phi$, which comes with a coupling constant that we denote by $\ell^{2}$. Note that its mass dimension is $[\ell]=-1$. Also, we assume $\ell^{2}>0$ -- this is necessary to ensure that vortex solutions are stable. The Lagrangian reduces to the usual Abelian Higgs model Lagrangian in the limit $\ell\rightarrow 0$. The derivative interaction has been choosen to preserve the original symmetries of the Abelian Higgs Model.
The symmetry groups of the Lagrangian are a global $U(1)$ symmetry:
\begin{equation}
\phi\rightarrow e^{i\alpha}\phi
\end{equation} with $\alpha$ being spatially independent. However, if the $\alpha$ is space-dependent, then the symmetry group is local $U(1)$, i.e.
\begin{equation}
\begin{split}
\phi&\rightarrow e^{i\alpha(x)}\phi\\
A_{\mu}&\rightarrow A_{\mu}-\frac{1}{e}\partial_{\mu}\alpha(x)\ .
\end{split}
\end{equation}
The Lagrangian has an imaginary mass term, i.e. $m^{2}>0$ and therefore the ground state spontaneously breaks the $U(1)$ gauge symmetry leading to massive photons. The potential has two stationary points
\begin{equation}
\label{vvv}
|\phi|=0\textrm{ and }|\phi|=\sqrt{\frac{m^{2}}{\lambda}}\equiv v\ ,
\end{equation}
and for $m^{2}>0$, the vacuum of the theory has a non-zero vacuum expectation value. This results in the photons becoming massive through the Higgs mechanism. Expanding around the vacuum expectation value (vev) we find the follows masses for the scalar and gauge fields
\begin{equation}
\begin{split}
m_{\phi}&=2m\\
m_{A}&=\frac{\sqrt{2}em}{\sqrt{\lambda}}\left (1+\frac{\ell^{2}m^{2}}{\lambda}\right )\ .
\end{split}
\end{equation}
Note that due to the derivative interaction which protects the local $U(1)$ symmetry, the photon mass is different from that of the standard Abelian Higgs model. In the limit $\ell\rightarrow 0$, we reproduce the photon mass in the Abelian Higgs model.
The equations of motion for $\phi$ and $A^{\mu}$ are respectively
\begin{equation}
\begin{split}
D_{\mu}D^{\mu}\phi-m^{2}\phi+\lambda|\phi|^{2}\phi+\ell^{2}D_{\mu}\left [(D^{\mu}\phi)\phi^{\dagger}\phi \right ]
=0\ ,
\end{split}
\end{equation}
and
\begin{equation}
\partial_{\mu}F^{\mu\nu}=-j^{\nu}\ ,
\end{equation}
where the electromagnetic current $j^{\nu}$ is
\begin{equation}
\begin{split}
j^{\nu}&=-ie\left [\phi^{\dagger}D^{\nu}\phi-(D^{\nu}\phi)^{\dagger}\phi \right ]\left[1+\ell^{2}\phi^{\dagger}\phi\right ]\ .
\end{split}
\end{equation}
\section{Single Vortex}
\label{mv1}
\noindent
Using the Lorentz gauge, $\partial_{\mu}A^{\mu}=0$, or equivalently the Landau gauge, since $A^{0}=0$, the equation of motion for $A^{j}$ is
\begin{equation}
-\partial_{i}\partial_{i}A^{j}=-j^{j}=ie\left [\phi^{\dagger}D^{j}\phi-(D^{j}\phi)^{\dagger}\phi \right ]\left[1+\ell^{2}\phi^{\dagger}\phi\right ]\ .
\end{equation}
For single vortex solutions, which are cylindrically symmetric, we will parameterize the vortices as follows 
\begin{equation}
\begin{split}
A_{r}&=A_{z}=0\\
A_{\theta}\equiv A&=\frac{n}{er}+\delta A\\
\phi(r,\theta)&=w(r) e^{in\theta}\ ,
\end{split}
\end{equation}
with the electromagnetic current taking the form
\begin{equation}
j_{\theta}=2 e^{2}|\phi|^{2}\left(A-\frac{n}{er}\right)\left[1+\ell^{2}|\phi|^{2}\right ]\ ,
\end{equation}
and the equation of motion for $A\equiv A_{\theta}$ becomes
\begin{equation}
\frac{\partial A}{\partial r^{2}}+\frac{1}{r}\frac{\partial A}{\partial r}-\frac{A}{r^{2}}=2 e^{2}|\phi|^{2}\left(A-\frac{n}{er}\right)\left[1+\ell^{2}|\phi|^{2}\right ]
\end{equation}
In a region, where $|\phi|=v\equiv\sqrt{\frac{m^{2}}{\lambda}}$, with $v$ being the vacuum expectation value, the equation can be solved using the following parametrization
\begin{equation}
A(r)=\frac{n}{er}+\delta A(r)\ ,
\end{equation}
where $\frac{n}{er}$ is the asymptotic behavior of the gauge field where the magnetic field is expected to be uniform and 
\begin{equation}
\begin{split}
\delta A(r)&\equiv \tilde{c}_{A} e^{-\sqrt{2}\tilde{v} er}\left ( \frac{1}{\sqrt{r}}+\mathcal{O}\left (\frac{1}{r^{3/2}} \right )\right )\ ,
\end{split}
\end{equation}
with $\tilde{c}_{A}$ being an undetermined constant and $\tilde{v}$ is defined as
\begin{equation}
\tilde{v}\equiv v\sqrt{1+\ell^{2}v^{2} }\ .
\end{equation}
Note that $\tilde{v}$ reduces to $v$ in the absence of derivative interactions, i.e. $\ell\rightarrow 0$.

We find the asymptotic properties of the complex scalar field $\phi$ using the parametrization
\begin{equation}
\phi(r,\theta)=w(r)e^{i\chi(\theta)} ,
\end{equation}
where $\chi=n\theta$. This choice guarantees that the electromagnetic current vanishes on the boundary leading to the flux quantization condition
\begin{equation}
\begin{split}
\int \vec{A}\cdot d\vec{l}=\int\ \vec{B}\cdot d\vec{S}=\frac{2n\pi}{e}\ ,
\end{split}
\end{equation}
where the line integral is performed at infinity and $d\vec{S}$ is the differential area.
The equation of motion in terms of $w(r)$ becomes
\begin{equation}
\begin{split}
&w''(r)+\frac{w'(r)}{r}-\left (\frac{n}{r}-eA(r)\right )^{2}w(r)+m^{2}w(r)-\lambda w(r)^{3}\\
&+\ell^{2} w(r)\left [w'(r)^{2}+\frac{w(r)}{r}\left (w'(r)+rw''(r)\right )\right ]\\
&-2\ell^{2}\left (\frac{n}{r}-2eA(r) \right )\left (\frac{n}{r}-eA(r) \right )w(r)^{3}=0\ .
\end{split}
\end{equation}
Next, rewriting $w(r)=v+\delta \phi(r)$, we get a linearized equation for $\delta \phi(r)$,
\begin{equation}
\begin{split}
&\left ( 1+\ell^{2} v^{2}\right )\delta\phi''(r)+\left ( 1+\ell^{2} v^{2}\right )\frac{\delta\phi'(r)}{r}-2v^{2}\lambda\delta\phi(r)\approx\ell^{2} 2ev^{3}n\frac{\delta A(r)}{r}
\end{split}
\end{equation}
It is straightforward to solve the linearized equation in the limit of large $r$, with final solution assuming the form
\begin{equation}
\begin{split}
\delta\phi(r)=\delta\phi_{h}(r)+\delta\phi_{p}(r)\ ,
\end{split}
\end{equation}
where the solution of the homogeneous equation is $\delta\phi_{h}(r)$ and the particular solution is $\delta\phi_{p}(r)$. We find that
\begin{equation}
\begin{split}
\lim_{r\rightarrow \infty}\delta\phi(r)_{h}&=c_{\phi}K_{1}\left (\overline{v}r \right )=c_{\phi}e^{-\overline{v}r}\left [\frac{1}{\sqrt{r}}+\mathcal{O}\left (\frac{1}{r^{3/2}} \right ) \right ]\ ,
\end{split}
\end{equation}
where $c_{\phi}$ is an arbitrary constant and $\overline{v}$ is defined as
\begin{equation}
\overline{v}=\sqrt{\frac{2v^{2}\lambda}{1+\ell^{2} v^{2}}}\ .
\end{equation}
The particular solution on the other hand has the form~\cite{Adhikari:2017oxb}
\begin{equation}
\lim_{r\rightarrow \infty}\delta\phi(r)_{p}=g_{\phi}\frac{e^{-\alpha r}}{r}\ ,
\end{equation}
where 
\begin{equation}
\alpha=2\sqrt{2}ev\left (1+\ell^{2}v^{2}  \right )\ .
\end{equation}
\begin{figure}[b]
\centerline{\includegraphics[width=9cm]{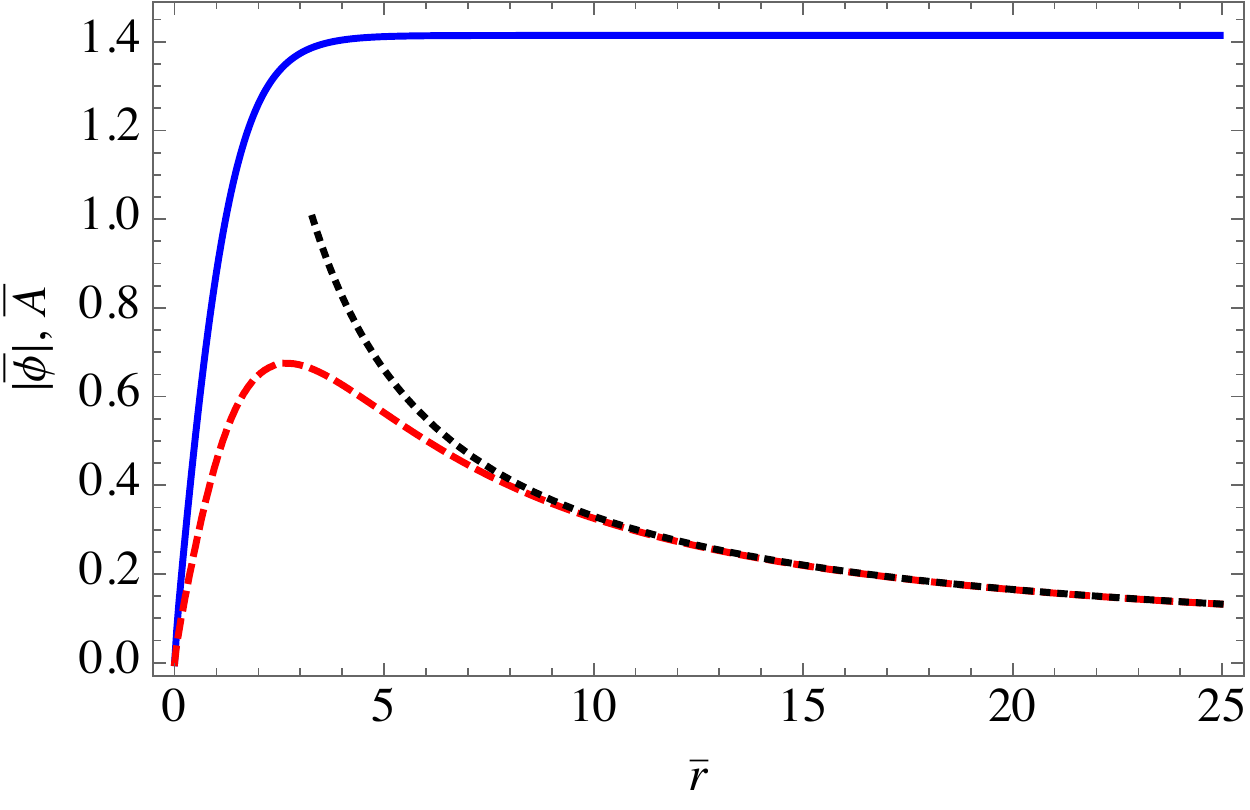}}
\caption{Single vortex solution for $\lambda=0.21$, $e^{2}=\frac{4\pi}{137}$, $\ell=0$, $m=1$, with $\bar{r}=mr$ and $|\bar{\phi}|=\frac{\phi}{m}$ and $\bar{A}=\frac{A}{m}$. The gauge field is shown using a dashed (red) curve and the scalar field using solid (blue). The dotted (black) curve represents $\frac{1}{e\bar{r}}$.}
\label{fig:gap1}   
\end{figure}
\begin{figure}[b]
\centerline{\includegraphics[width=9cm]{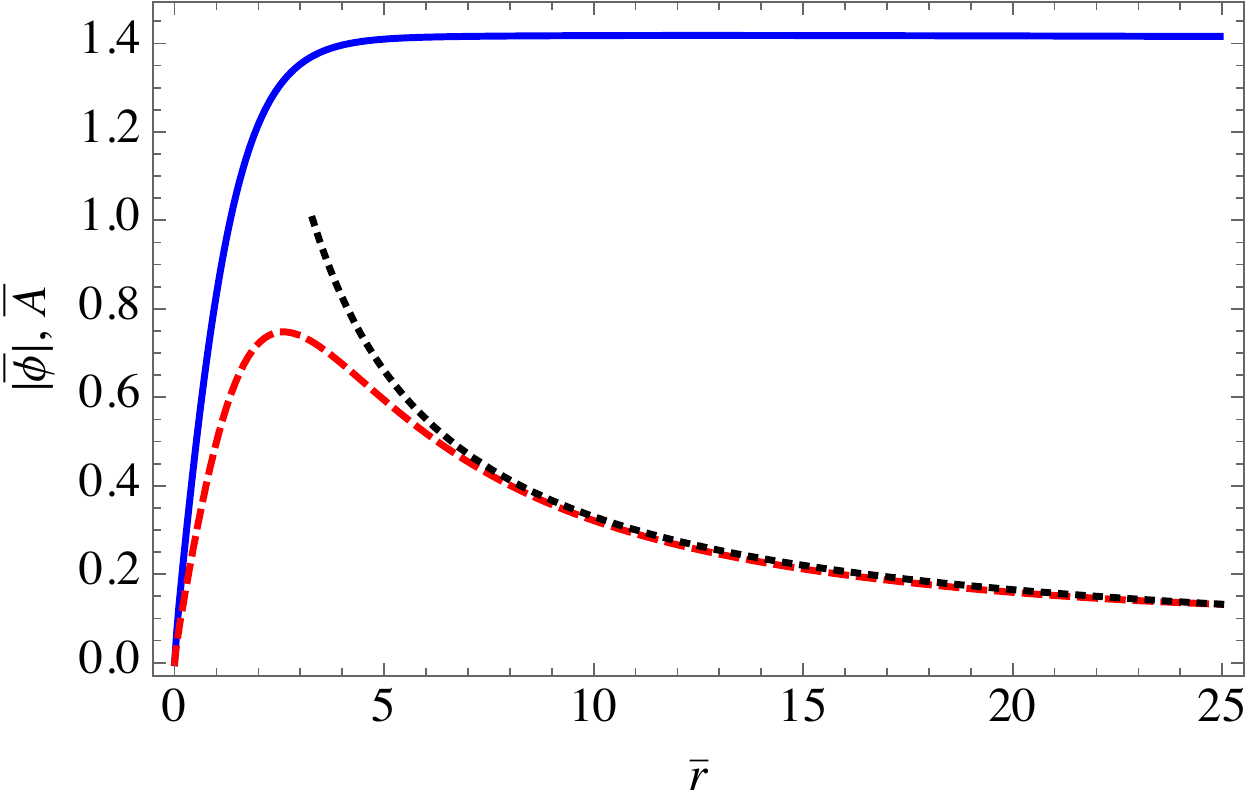}}
\caption{Single vortex solution for $\lambda=0.21$, $e^{2}=\frac{4\pi}{137}$, $\ell=\frac{1}{2}$, $m=1$, with $\bar{r}=mr$ and $|\bar{\phi}|=\frac{\phi}{m}$ and $\bar{A}=\frac{A}{m}$. The gauge field is shown using a dashed (red) curve and the scalar field using solid (blue). The dotted (black) curve represents $\frac{1}{e\bar{r}}$.}
\label{fig:gap2}  
\end{figure}
In Figs.~\ref{fig:gap1} and \ref{fig:gap2}, we plot single vortex solutions in a uniform magnetic field with flux $\frac{2\pi}{e}$. The solid (blue) curve is the pion field, the red (dashed) curve is the gauge field and the black (dotted) line indicates a uniform magnetic field, i.e. $A=\frac{1}{er}$. Firstly, note that away from the vortex core ($r=0$), the scalar field asymptotes to its vev and the magnetic field is uniform. 
Fig.~\ref{fig:gap1} is a single vortex in the standard Abelian Higgs Model, i.e. $\ell=0$ and Fig.~\ref{fig:gap2} is a single vortex with the derivative interaction turned on. The gauge field asymptotes to $\frac{1}{er}$ faster than when $\ell=0$. This is as anticipated since the photon mass in the presence of derivative interactions is larger than in the standard Abelian Higgs model.
\section{Vortex Lattice and Condensation Energy}
\label{condensation}
In this section, we find the vortex lattice solutions and their corresponding free energies near the upper critical field, $B_{c2}$ and find the condensation (free) energy of these lattice. We define the condensation energy as
\begin{equation}
\begin{split}
\mathcal{E}=\langle\mathcal{H}\rangle-\frac{1}{2}B_{\rm ext}^{2}\ ,
\end{split}
\end{equation}
where $\frac{1}{2}B_{\rm ext}^{2}$ is the free energy of the normal vacuum in the presence of a uniform magnetic field $B_{\rm ext}$ and $\langle\mathcal{H}\rangle$ is the expectation value of the time-dependent Hamiltonian (density), $\mathcal{H}$, which is defined as
\begin{equation}
\begin{split}
\langle \mathcal{H}\rangle=\frac{1}{A_{\perp}}\int\ dxdy\ \mathcal{H}\ ,
\end{split}
\end{equation}
where $x$ and $y$ are the coordinates of the transverse plane and we assume that the external magnetic field $B_{\rm ext}$ is pointing in the positive z-direction. The condensation (free) energy of the vortex lattice is defined relative to the normal vacuum in a uniform magnetic field.
We use the method of successive approximation originally used by Abrikosov in Ref.~\cite{Abrikosov:1956sx}. We will solve the equations of motion near the upper critical point, $B_{c2}=\frac{m^{2}}{e}$, which is identical to the critical field in the standard version of the Abelian Higgs Model. We will work just below the critical field, $B_{c2}$, i.e.
\begin{equation}
\frac{B_{c2}-B_{\rm ext}}{B_{c2}}\ll 1,
\end{equation}
a region where the condensate is expected to be small,
\begin{equation}
\frac{|\phi|}{v}\ll 1\ ,
\end{equation}
with $v$ being the vacuum expectation value in the condensed phase, i.e. $v>0$, defined in Eq.~\ref{vvv}. 

The method of ``successive approximations" is effectively the following perturbative expansion~\footnote{This expansion is commonly used to study non-linear harmonic oscillators in classical mechanics.}
\begin{equation}
\begin{split}
B&=B_{0}+\epsilon \delta B+\cdots\\
\phi&=\epsilon \phi_{0}+\epsilon^{3} \delta \phi+\cdots\\
\end{split}
\end{equation}
Similarly, the expansion for the gauge field, $A_{i}$, that is consistent with the expansion for the magnetic field $B$, is 
\begin{equation}
\begin{split}
A_{x}&=A_{x0}+\epsilon\delta A_{x}\\
A_{y}&=A_{y0}+\epsilon\delta A_{y}\\
A_{z}&=0\ .
\end{split}
\end{equation}
The $\mathcal{O}(\epsilon)$ equation of motion is
\begin{equation}
\begin{split}
\left [-(\partial_{i}+ieA_{0i})^{2}-m^{2}\right]\phi_{0}=0\ ,
\end{split}
\end{equation}
which we rewrite in terms of 
\begin{equation}
\begin{split}
\Pi_{0x}&=-i\partial_{x}-eA_{0x}\\
\Pi_{0y}&=-i\partial_{y}-eA_{0y}\ .
\end{split}
\end{equation}
They satisfy the commutation relation $[\Pi_{x},\Pi_{y}]=ieB_{0}$, where $B_{z}$ is the magnetic field in the z-direction. Further, defining
\begin{equation}
\begin{split}
\Pi_{0}&=\Pi_{0x}+i\Pi_{0y}\\
\Pi^{\dagger}_{0}&=\Pi_{0x}-i\Pi_{0y}\ ,
\end{split}
\end{equation}
we can rewrite the equation of motion as 
\begin{equation}
\begin{split}
(\Pi_{0}^{\dagger}\Pi_{0}+eB_{0}-m^{2})\phi_{0}=0 .
\end{split}
\end{equation}
Since $[\Pi,\Pi^{\dagger}]=2eB_{0}$, which now has a structure similar to that of the simple harmonic oscillator, we know that the dispersion relation is
\begin{equation}
\begin{split}
(2n+1)eB_{0}-m^{2}=0\implies B_{0}=\frac{m^{2}}{(2n+1)e}\ .
\end{split}
\end{equation}
The maximum magnetic field, $B_{0}$, is therefore,
\begin{equation}
\begin{split}
B_{0}=\frac{m^{2}}{e}\equiv B_{c2}\ .
\end{split}
\end{equation}
Below this field and above the first critical field $B_{c1}$, a phase of magnet vortices form in type-II superconductors. $B_{c2}$ is the largest magnetic field that can be sustained by magnetic vortices, beyond which the repulsive interactions between the vortices becomes large enough such that it becomes energetically favorable to condense into the normal vacuum. Just below the the critical field $B_{c2}$, the vortex solutions  satisfy  $\Pi_{0}\phi_{0}=0$. In terms of the complex variables $z$ and $\bar{z}$, which we define below 
\begin{equation}
\begin{split}
z&=x+iy,\ \overline{z}=x-iy\\
\partial&=\frac{1}{2}(\partial_{x}-i\partial_{y}),\ \overline{\partial}=\frac{1}{2}(\partial_{x}+i\partial_{y})\\
\end{split}
\end{equation}
the equation of motion at leading order is
\begin{equation}
\left (2\overline{\partial}+\frac{eB_{0}}{2}z \right )\phi_{0}=0\ ,
\end{equation}
Working in the symmetric gauge with
\begin{equation}
\begin{split}
A_{0x}=-\frac{B_{0}}{2}y,\ A_{0y}=\frac{B_{0}}{2}x\ , 
\end{split}
\end{equation}
the ground state is a sum over the lowest Landau levels~\cite{Chernodub:2011gs}
\begin{equation}
\label{phi0}
\begin{split}
\phi_{0}&=\sum_{n=-\infty}^{\infty}C_{n}\phi_{n}(\nu,z,\bar{z})\\
\phi_{n}(\nu,z,\bar{z})&=e^{-\pi\nu^{2}n^{2}-\frac{\pi}{2L_{B_{0}}}\left (|z|^{2}+z^{2} \right)+\frac{2\pi}{L_{B_{0}}}\nu n z}\ ,
\end{split}
\end{equation}
where $L_{B_{0}}$ is the ``reduced magnetic length", $L_{B_{0}}\equiv\sqrt{\frac{2\pi}{eB_{0}}}$, $n$ is an integer and $\nu$ is a variational parameter. Since there are an infinite number of possible solutions, i.e. choice of a set of $C_{n}$, we proceed by imposing a periodicity constraint such that $C_{n}=C_{n+N}$. $N$ determines the lattice structure. (For example, $N=1$ is a square lattice, $N=2$ is a hexagonal (triangular) lattice and $N=3$ is a parallelogramic lattice. See Ref.~\cite{Chernodub:2011gs,saint1969type} for further details.)

In order to find the lattice structure and the corresponding condensation energy, we need to work at next to leading order, i.e. $\mathcal{O}(\epsilon^{3})$, in our expansion. Using the equation of motion for $F^{\mu\nu}$, where the current $j^{\nu}$ is determined by $\phi_{0}$, we get
\begin{equation}
\overline{\partial} B=\frac{j^{2}-i j^{1}}{2}=-e\overline{\partial}(\phi_{0}^{\dagger}\phi_{0})\left [1+\ell^{2}\phi_{0}^{\dagger}\phi_{0} \right ]\ .
\end{equation}
Note that since $B=B_{c2}$ at $\mathcal{O}(\epsilon)$, we have $\overline{\partial}B=\overline{\partial}\delta B$, we can solve the above equation using the following ansatz:
\begin{equation}
\label{BBB}
B=B_{\rm ext}-e(|\phi_{0}|^{2}-c\langle |\phi_{0}|^{2}\rangle)-\frac{e}{2}\ell^{2}(|\phi_{0}|^{4}-c\langle |\phi_{0}|^{4}\rangle)\ ,
\end{equation}
where $c=1$ is the normalization chosen in Ref.~\cite{Chernodub:2014rya} and $c=0$ is the normalization used in Ref.~\cite{Abrikosov:1956sx}. We choose the normalization with $c=1$ since it guarantees the conservation of magnetic flux~\cite{Chernodub:2014rya}. 

Next, we consider the equation of motion for the scalar field, $\phi$, at $\mathcal{O}(\epsilon^{3})$~\cite{SDreview}:
\begin{equation}
\begin{split}
(\Pi_{0x}^{2}+\Pi_{0y}^{2}-m^{2})\phi_{1}-\frac{e}{2}(\Pi_{0x}\delta A_{x}+\Pi_{0y}\delta A_{y}+\delta A_{x}\Pi_{0x}+\delta A_{y}\Pi_{0y})\phi_{0}+\lambda|\phi_{0}|^{2}\phi_{0}=0
\end{split}
\end{equation}
Rewriting in terms of $\Pi_{0}$ and $\Pi_{0}^{\dagger}$, we get
\begin{equation}
\begin{split}
(\Pi_{0}^{\dagger}\Pi_{0}+eB_{c2}-m^{2})\phi_{1}-\frac{e}{2}(\Pi_{0}\delta \bar{A}+\delta \bar{A}\Pi_{0}+\Pi_{0}^{\dagger}\delta A+\delta A \Pi_{0}^{\dagger})\phi_{0}+\lambda |\phi_{0}|^{2}\phi_{0}=0\ ,
\end{split}
\end{equation} 
where $\delta A=\delta A_{x}+i\delta A_{y}$ and $\delta \bar{A}=\delta A_{x}-i\delta A_{y}$. Multiplying on the left by $\phi_{0}^{\dagger}$, using the fact that $\phi_{0}^{\dagger}\Pi_{0}^{\dagger}=0$, $B_{c2}=\frac{m^{2}}{e}$ and averaging over the transverse plane, we get the condition
\begin{equation}
\begin{split}
\frac{1}{A_{\perp}}\int dx dy \left [-\phi_{0}^{\dagger}\frac{e}{2}(\Pi_{0}\delta\bar{A}+\delta\bar{A}\Pi_{0}+\Pi_{0}^{\dagger}\delta A+\delta A\Pi_{0}^{\dagger})\phi_{0}+\lambda |\phi_{0}|^{4} \right ]=0\ .
\end{split}
\end{equation}
Since $\Pi_{0}\phi_{0}=0$ (and consequently $\phi_{0}^{\dagger}\Pi_{0}^{\dagger}=0$), it is useful to move all the $\Pi_{0}$s to the right and $\Pi_{0}^{\dagger}$s to the left using the commutator identities
\begin{equation}
\begin{split}
\Pi_{0}\delta \bar{A}&=[\Pi_{0},\delta\bar{A}]+\delta\bar{A}\Pi_{0}=2i\bar{\partial}\delta\bar{A}+\delta\bar{A}\Pi_{0}\\
\delta A\Pi_{0}^{\dagger}&=\Pi_{0}^{\dagger}\delta A-[\Pi_{0}^{\dagger},\delta A]=\Pi_{0}^{\dagger} \delta A+2i\partial \delta A\ .
\end{split}
\end{equation}
The result after averaging over the transverse plane is
\begin{equation}
\begin{split}
\langle -ie (\partial\delta A-\bar{\partial}\delta \bar{A})|\phi_{0}|^{2}+\lambda\langle |\phi_{0}|^{2}\rangle=0\ .
\end{split}
\end{equation}
Since
\begin{equation}
\begin{split}
-i(\partial \delta A-\bar{\partial}\delta \bar{A})=\delta B=B-B_{c2}\ .
\end{split}
\end{equation}
we get
\begin{equation}
\label{R}
\mathcal{R}\equiv \frac{\langle |\phi_{0}|^{4}\rangle}{\langle|\phi_{0}|^{2} \rangle}=\frac{e(B_{c2}-B_{\rm ext})-eB_{1}}{\lambda-e^{2}+2m^{2}\ell^{2}}\ ,
\end{equation}
where 
\begin{equation}
B_{1}=\langle|\phi_{0}|^{2}\rangle+\frac{1}{2}\ell^{2}\langle |\phi_{0}|^{4}\rangle\ .
\end{equation}
Finally, we can write down the expectation value of the time-independent Hamiltonian (i.e. free energy) using the expression for $\mathcal{R}$ above, the equations of motion $\phi_{0}$, which can be used to get rid of all terms containing derivatives and $B$ found in Eq.~\ref{BBB}
\begin{equation}
\label{freeenergy}
\begin{split}
\langle\mathcal{H}\rangle&=\frac{1}{2}\langle B^{2}\rangle-\left ( \frac{\lambda}{2}+m^{2}\ell^{2}\right )\langle|\phi_{0}|^{4}\rangle+\mathcal{O}(|\phi_{0}|^{6})\\
&=\frac{1}{2}B_{\rm ext}^{2}-e(B_{c2}-B_{\rm ext})\langle|\phi_{0}|^{2}\rangle+\frac{e^{2}}{2} \langle |\phi_{0}|^{2}\rangle^{2}\\
&+\frac{1}{2}\left (\lambda-e^{2}+2m^{2}\ell^{2}\right )\langle|\phi_{0}|^{4}\rangle+\mathcal{O}(|\phi_{0}|^{6})\ ,
\end{split}
\end{equation}
where we are ignoring all contributions containing more than four fields, which is valid if $\phi_{0}\ll v$. 
The (Gibbs free) energy can be minimized using the replacement
\begin{equation}
\langle|\phi_{0}^{4}| \rangle\equiv \beta_{A}{\langle|\phi_{0}^{2}| \rangle^{2} }\ ,
\end{equation}
where $\beta_{A}\ge 1$ is the Abrikosov ratio~\cite{Abrikosov:1956sx,abrikosov1957magnetic,abrikosov1988fundamentals} and subsequently minimizing with respect to $\langle|\phi_{0}|^{2} \rangle$, followed by $\beta_{A}$. We find that up to the order to which we are working the free energy is minimized by 
\begin{equation}
\langle |\phi_{0}|^{2}\rangle=\frac{e(B_{c2}-B_{\rm ext})}{\beta_{A}\left (\lambda-e^{2}+2m^{2}\ell^{2}\right)+e^{2}}+\cdots
\end{equation}
and the expectation value of the Hamiltonian for the lattice (free energy averaged over the transverse plane) given by
\begin{equation}
\langle \mathcal{H}\rangle=\frac{1}{2}B_{\rm ext}^{2}-\frac{e^{2}(B_{c2}-B_{\rm ext})^{2}}{2e^{2}+2\beta_{A}\left (\lambda-e^{2}+2m^{2}\ell^{2} \right )}+\mathcal{O}((B_{c2}-B_{\rm ext})^{4})\ .
\end{equation}
Note that the free energy reduces to the standard result in the limit $\ell\rightarrow 0$. Furthermore, we see from the above expression that the free energy minimized by a vortex lattice with the smallest value of the Abrikosov ratio. The hexagonal lattice assume the smallest value of $\beta_{A}=1.1596\dots$, which has a corresponding value of $\nu=\frac{\sqrt[4]{3}}{\sqrt{2}}$ for the hexagonal lattice. As expected the free energy in the presence of the derivative interaction with $\ell^{2}>0$ is higher than without.
\subsection{Vortex Lattice in an external field $B_{\rm ext}$}
The Abrikosov ratio that minimizes the free energy is that of a periodic hexagonal lattice with $C_{0}=C$, $C_{1}=iC$ and $C_{n}=C_{n+2}$. However, the solution of Eq.~(\ref{phi0}) with the hexagonal periodicity constraints only obeys the flux quantization condition 
\begin{equation}
B_{\rm ext}\ \left |\vec{d}_{1}\times\vec{d}_{2}\right |=\frac{2\pi n}{e}\ ,
\end{equation}
at the critical field, i.e. $B_{\rm ext}=B_{c2}$ with $n=1$ in each unit cell. Note that $\vec{d}_{1}=\frac{L_{B}}{\nu}(0,1)$ and $\vec{d}_{2}=\frac{L_{B}}{\nu}\left(\frac{\sqrt{3}}{2},\frac{1}{2}\right)$ are the lattice vectors of the periodic lattice with the reduced magnetic length $L_{B}=\sqrt{\frac{2\pi}{B}}$. In order to ensure the appropriate flux quantization condition, we make the change $L_{B0}\rightarrow L_{B}$ in Eq.~(\ref{phi0}) with the guarantee that the vortex condensation energy we found remains unaffected since
\begin{equation}
\begin{split}
\langle\phi_{0}^{\dagger}\phi_{0} \rangle&=\frac{|C|^{2}}{\sqrt{2}|\nu|}=\frac{|C|^{2}}{\sqrt[4]{3}}\\
\langle(\phi_{0}^{\dagger}\phi_{0})^{2} \rangle&=\frac{|C|^{4}}{4|\nu|}\left [\theta_{3}(0,e^{-\pi\nu^{2}})^{2}+2\theta_{3}(\pi/2,e^{-\pi\nu^{2}})\theta_{3}(0,e^{-\pi\nu^{2}})-\theta_{3}(-\pi/2,e^{-\pi\nu^{2}})^{2} \right ]\\\
&=\frac{\beta_{A}|C|^{4}}{\sqrt{3}}
\end{split}
\end{equation}
where $\theta_{3}$ is a Jacobi Theta function. The condensation energy is a function of $\beta_{A}$ and $\langle\phi_{0}^{\dagger}\phi_{0} \rangle$, the latter of which only depends on the quantity $C$ and the Abrikosov ration ($\beta_{A}$) is determined by the geometry of the vortex lattices.

\begin{figure*}[t!]
\centering
\includegraphics[width=0.49\textwidth]{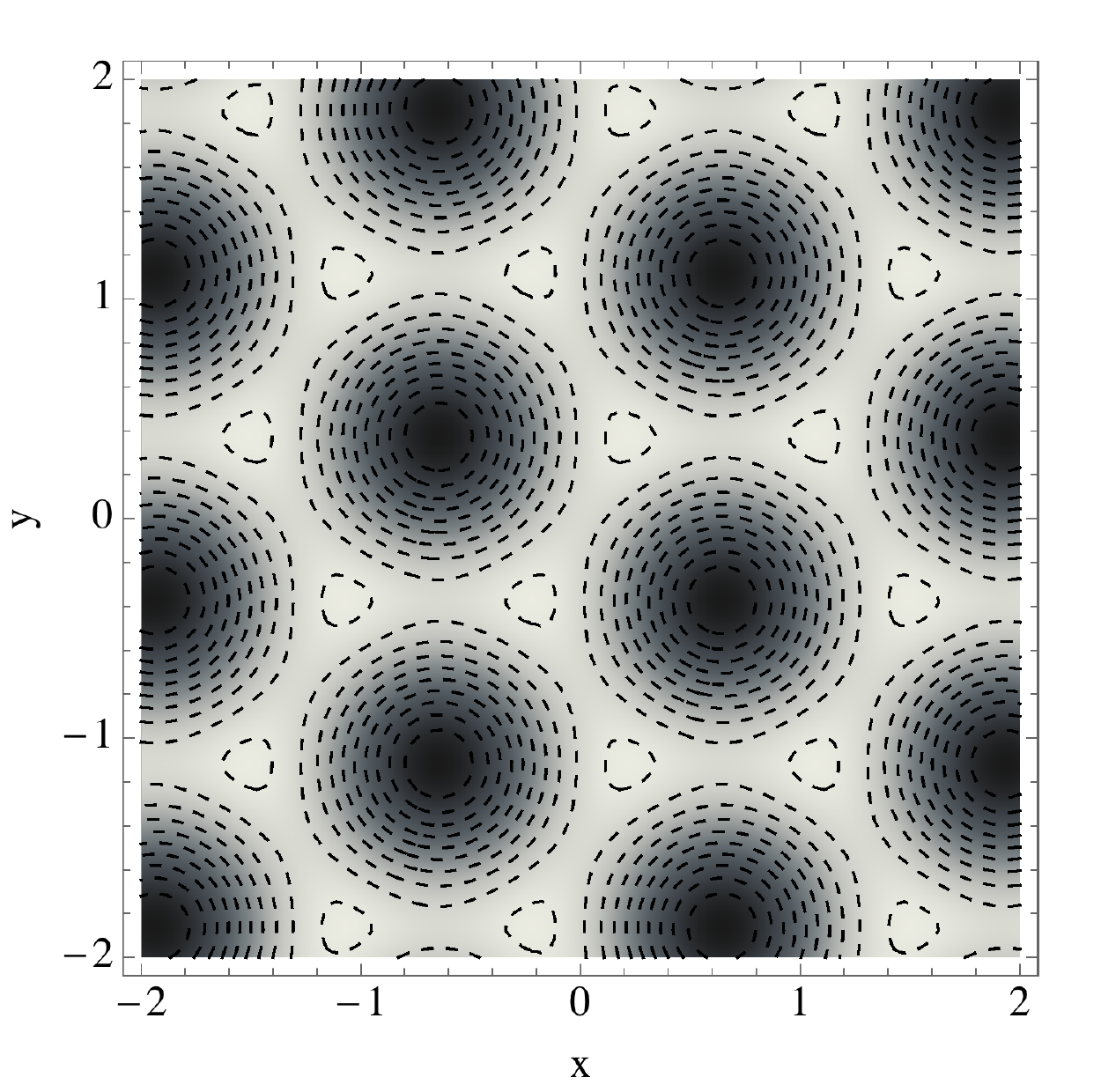}\,\,\,\,\,
\includegraphics[width=0.49\textwidth]{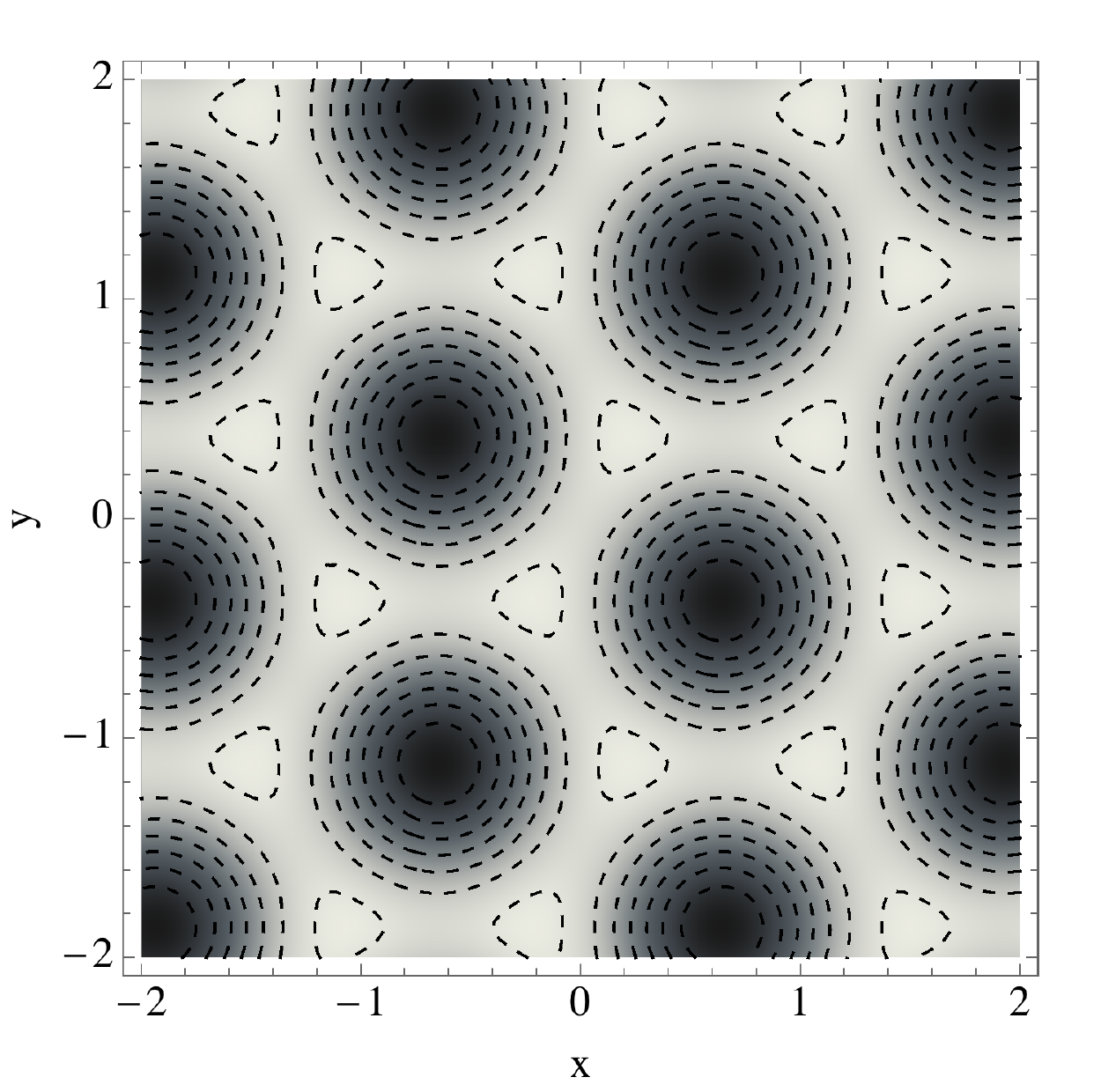}
\caption{Vortex Lattice Plot of ($|\phi^{\dagger}\phi|$) on the traverse-plane ($x-y$) of size $2\times 2$. We use $m=1$ to set the scale, $\lambda=1$, $e=\sqrt{\frac{4\pi}{137}}$ and from left to right $m\ell^{2}=0,0.2,$ respectively. Note that the color shades are normalized differently for each plot but the contour lines have the same ``normalization" across all the plots. We note that from left to right $|C|=0.107213, 0.0904461$.} 
\label{fig:vlattice}     
\end{figure*}

Now, we can finally write the vortex lattice solution in terms of $L_{B}$ and the periodic sum for the hexagonal lattice can be rewritten in terms of the third Jacobi (elliptic) theta function, $\theta_{3}$ as follows
\begin{equation}
\label{phigs}
\begin{split}
\phi_{0}(x,y)&=\frac{C}{2\nu}\left [ e^{-\frac{\pi y(-i x+y)}{L_{B}^{2}}}\theta_{3}\left(\frac{-\pi(x+iy)}{2L_{B}\nu},e^{-\frac{\pi}{4\nu^{2}}} \right)\right.\\
&\left.+ie^{-\frac{\pi y(-ix+y)}{L_{B}^{2}}}\theta_{3}\left(\frac{-\pi (x+i y-L_{B}\nu)}{2L_{B}\nu},e^{-\frac{\pi}{4\nu^{2}}} \right)\right ]\ ,
\end{split}
\end{equation}
where $\nu=\frac{\sqrt[4]{3}}{\sqrt{2}}$ for the hexagonal lattice that minimizes the free energy.
Finally, we determine $C$ up to an arbitrary phase constant by minimizing the condensation energy. We get
\begin{equation}
\begin{split}
\label{C}
&|C|=\sqrt{\frac{\sqrt[4]{3}e(B_{c2}-B_{\rm ext})}{\beta_{A}\left (\lambda-e^{2}+2m^{2}\ell^{2}\right )+e^{2}}}\ ,{\rm\ for\ }B_{\rm ext}\le B_{c2}\\
&|C|=0\ ,{\rm\ for\ }B_{\rm ext}\ge B_{c2}\ .
\end{split}
\end{equation}
Using the above result for $C$, we plot the vortex lattice. The vortex lattices as shown ($\phi^{\dagger}\phi$ plotted) in Fig.~\ref{fig:vlattice}. We choose $m=1$ such that distance scales, such that $x$ and $y$ shown in the plots are effectively measured in units of $m$. Also, we choose $\lambda=1$. It is evident from the plots that changing $\ell^{2}$ does not affect the density but only the gradients and the maximum value of $|C|_{\rm max}$. With increasing values of $\ell^{2}$, the gradient in $\phi$ becomes energetically expensive and therefore $\phi_{0}^{\dagger}\phi_{0}$ becomes flatter as the plots suggest. But note that the density of vortices is unaffected by a change in $\ell^{2}$. The density only depends on the external magnetic field ($B_{\rm ext}$), which also affects $|C|$.

\section{Conclusion}
\label{conclusion}
In this paper, we have found the asymptotic properties of single vortex solutions in the presence of an additional derivative interaction. Furthermore, we have also found the condensation energy and the corresponding vortex lattice structure associated with vortices that form near the upper critical field $B_{c2}$. One of the goals of this paper was to lay the groundwork to study vortex lattice solutions and the corresponding condensation energy in finite isospin chiral perturbation theory, where pions are Goldstone bosons and interact via momentum-dependent derivative interactions -- the possibility of  vortex condensation was first discussed in Ref.~\cite{Adhikari:2015wva}. A detailed analysis of vortex lattice solutions in finite isospin chiral perturbation theory will be done in a separate publication~\cite{Adhikari:new}.
\bibliographystyle{elsarticle-num} 
\end{document}